\documentclass[MSNbibl,nameyear,dvips]{arxstspdf}
\usepackage{flushend}
\usepackage{stfloats}

\volume{26}
\issue{1}
\pubyear{2011}
\firstpage{102}
\lastpage{115}
\doi{10.1214/10-STS351}

\makeatletter

\def\bsuffix #1{#1}

\newproclaim{definition}{Definition}
\makeatother

\begin{document}
\begin{frontmatter}

\title{A Short History of Markov Chain Monte Carlo: Subjective
Recollections from
Incomplete Data\thanksref{T1}}
\runtitle{A Short History of Markov
Chain Monte Carlo}
\thankstext{T1}{A shorter version of this paper appears
as Chapter 1 in \textit{Handbook of Markov Chain Monte Carlo} (2011),
edited by Steve Brooks, Andrew Gelman, Galin Jones and Xiao-Li Meng.
Chapman \& Hall/CRC Handbooks of Modern Statistical Methods, Boca
Raton, Florida.}

\begin{aug}
\author{\fnms{Christian} \snm{Robert}\ead[label=e1]{xian@ceremade.dauphine.fr}}
\and
\author{\fnms{George} \snm{Casella}\corref{}\ead[label=e2]{casella@ufl.edu}}
\runauthor{C. Robert and G. Casella}

\affiliation{Universit\'{e} Paris
Dauphine and CREST, INSEE and University of Florida}
\vspace*{-12pt}
\dedicated{This paper is dedicated to the memory of our friend Julian Besag,
a~giant in the field of MCMC.}

\address{Christian Robert is Professor of Statistics, CEREMADE,
Universit\'{e} Paris
Dauphine, 75785 Paris cedex 16, France \printead{e1}. George Casella
is Distinguished Professor, Department of Statistics, University of Florida,
Gainesville, Florida 32611, USA \printead{e2}.}

\end{aug}

%
\begin{abstract}
We attempt to trace
the history and development of Markov chain Monte Carlo (MCMC) from its
early inception in the late
1940s through its use today. We see how the earlier stages of Monte
Carlo (MC, not MCMC)
research have led to the algorithms currently in use.
More importantly, we see how the development of this methodology has
not only changed our solutions
to problems, but has changed the way we think about problems.
\end{abstract}

%
\begin{keyword}
\kwd{Gibbs sampling}
\kwd{Metropolis--Hasting algorithm}
\kwd{hierarchical models}
\kwd{Bayesian methods}.
\end{keyword}

\end{frontmatter}

\section{Introduction}

Markov chain Monte Carlo (MCMC) methods ha\-ve been around for almost as long
as Monte Carlo techniques, even though their impact on Statistics has
not been
truly felt until the very early 1990s, except in the specialized fields
of Spatial
Statistics and Image Analysis, where those methods appeared earlier.
The emergence
of Markov based techniques in Physics
is a~story that remains\vadjust{\goodbreak} untold within this survey (see Landau and Binder, \citeyear{LandauBinder2005}).
Also, we will not enter into a~description of MCMC techniques. 
A comprehensive treatment of MCMC techniques, with further references,
can be found in Robert and Casella (\citeyear{robertcasella2004}).

We will distinguish between the introduction of Metropolis--Hastings
based algorithms and
those related to Gibbs sampling, since they each stem from radically
different origins, even
though their mathematical justification via Markov chain theory is~the same.
Tracing the development of Monte Carlo methods, we will also briefly
mention what we might call the
``second-generation MCMC revolution.'' \mbox{Starting} in the mid-to-late
1990s, this includes the development of
particle filters, reversible jump and perfect sampling, and concludes
with more current work on population
or sequential Monte Carlo and regeneration and the computing of
``honest'' standard errors.

\setcounter{footnote}{1}
As mentioned above, the realization that Markov chains could be used
in a wide variety of situations only came (to mainstream statisticians)
with Gelfand and Smith (\citeyear{GelfandSmith1990}),
despite earlier publications in the statistical literature like Hastings (\citeyear{Hastings1970}),
Geman and Geman (\citeyear{GemanGeman1984}) and Tanner and Wong (\citeyear{TannerWong1987}).
Several reasons can be advanced: lack of computing machinery (think of\vadjust{\goodbreak}
the computers of 1970!), or background on Markov chains, or hesitation
to trust in the practicality of the method. It thus required
visionary researchers like Gelfand and Smith to convince the community,
supported by
papers that demonstrated, through a series of applications, that the
method was easy to
understand, easy to implement and practical
(Gelfand et~al., \citeyear{gelfandhillsracinepoonsmith1990};
Gelfand, Smith and Lee, \citeyear{gelfandsmithlee1992};
Smith and Gelfand, \citeyear{smithgelfand1992}; Wakefield et~al., \citeyear{wakefieldsmithracinepoongelfand1994}).
The rapid emergence
of the dedicated BUGS (Bayesian inference Using Gibbs Sampling)
software as early as
1991, when a~paper on BUGS was presented at the Valencia meeting, was
another compelling argument for
adopting, at large, MCMC algorithms.\footnote{Historically speaking,
the development of BUGS initiated from
Geman and Geman (\citeyear{GemanGeman1984}) and Pearl (\citeyear{Pearl1987}), in accord with the
developments in the artificial intelligence community,
and it predates Gelfand and Smith (\citeyear{GelfandSmith1990}).}

\section{Before the Revolution}

Monte Carlo methods were born in Los Alamos, New Mexico during World
War II, eventually resulting in the Metropolis algorithm
in the early 1950s. While Monte Carlo methods were in use by that time,
MCMC was brought closer to statistical
practicality by the work of Hastings in the 1970s.

What can be reasonably seen as the first MCMC algorithm is what we now
call the Metropolis algo\-rithm,
published by Metropolis et~al. (\citeyear{MetropolisRosenbluthRosenbluthTellerTeller1953}). It emanates
from the same group of scientists who produced the Monte Carlo method,
namely, the research scientists of
Los Alamos,
mostly physicists working on mathematical physics and the atomic bomb.%

MCMC algorithms therefore date back to the same time as the development
of regular (MC only) Monte Carlo methods, which are usually traced to
Ulam and von Neumann in the late 1940s. Stanislaw Ulam associates the
original idea with an intractable combinatorial computation he
attempted in 1946 (calculating the probability of winning at the card
game ``solitaire''). This idea was
enthusiastically adopted by John von Neumann for implementation with
direct applications to neutron diffusion, the name\break ``Monte Carlo''
being suggested by Nicholas Metropo\-lis.
(Eckhardt, \citeyear{Eckhardt1987}, describes these early Monte~Car\-lo
developments, and Hitchcock, \citeyear{hitch2003}, gives a~brief history of the
Metropolis algorithm.)

These occurrences very closely coincide with
the appearance of the very first computer, the ENIAC, which came to
life in February
1946, after three years of construction. The Monte Carlo\vadjust{\goodbreak} method was set
up by von
Neumann, who was using it on thermo\-nuclear and fission problems as
early as 1947. At the same
time, that is, 1947, Ulam and von Neumann~in\-vented inversion and
accept-reject techniques (also recounted in Eckhardt, \citeyear{Eckhardt1987}) to
simulate from nonu\-niform distributions. Without computers, a
rudimen\-tary version
invented by Fermi in the 1930s did not get any recognition (Metropolis, \citeyear{metropolis1987}).
Note~also~that, as early as 1949, a symposium on Monte~Car\-lo was
supported by Rand, NBS and
the Oak Ridge~labora\-tory and that Metropolis and Ulam (\citeyear{metropolisulam1949}) published
the very first paper about the Monte Carlo~method.

\setattribute{citeyear}{font}{\sffamily}
\subsection{\texorpdfstring{The Metropolis et al. (\citeyear{MetropolisRosenbluthRosenbluthTellerTeller1953}) Paper}%
{The Metropolis et al. (1953) Paper}}
\setattribute{citeyear}{font}{\relax}

The first MCMC algorithm is associated with a~se\-cond computer,
called MANIAC, built\footnote{MANIAC stands for \textit{Mathematical
Analyzer, Numerical Integrator and Computer.}} in Los Ala\-mos under the
direction of Metropolis in early 1952. Both a physicist and a mathematician,
Nicolas Me\-tropolis, who died in Los Alamos in 1999, came to this~pla\-ce
in April 1943. The other members of the team~also
came to Los Alamos during those years, including the controversial
Teller. As early as 1942,
he~became obsessed with the hydrogen (H) bomb, which~he eventually
managed to design with Stanislaw Ulam, using the better
computer facilities in the early 1950s.

Published in June 1953 in the \textit{Journal of Chemical Physics}, the primary
focus of Metropolis et~al. (\citeyear{MetropolisRosenbluthRosenbluthTellerTeller1953}) is the
computation of integrals of the form
\begin{eqnarray*}
\mathfrak{I} &=& {\int F(\theta) \exp\{ -E(\theta)/kT \}\,
{d}\theta}\\
&&\Big/{\int\exp\{ -E(\theta)/kT \}\,{d}\theta},
\end{eqnarray*}
on $\mathbb{R}^{2N}$, $\theta$ denoting a set of $N$ particles on
$\mathbb{R}^{2}$,
with the energy $E$ being defined as
\[
E(\theta) = \frac{1}{2}\sum_{i=1}^N\sum_{\stackrel{j=1}{j\ne i}}^N
V(d_{ij}),
\]
where $V$ is a potential function and $d_{ij}$ the Euclidean distance between
particles $i$ and $j$ in $\theta$. The \textit{Boltzmann distribution}
$\exp\{ -E(\theta)/kT \}$ is parameterized by
the \textit{temperature} $T$, $k$ being the Boltzmann constant, with a~normalization factor
\[
Z(T) = \int\exp\{ -E(\theta)/kT \}\,{d}\theta,
\]
that is not available in closed form, except in trivial cases.
Since $\theta$ is a $2N$-dimensional vector,\vadjust{\goodbreak} numerical integration
is impossible. Given the large dimension of the problem, even standard
Monte Carlo techniques fail to
correctly approximate $\mathfrak{I}$, since\break
$\exp\{ -E(\theta)/kT \}$ is very small for most realizations of the
random configurations of the
particle system (uniformly in the $2N$ square). In order to improve the
efficiency of
the Monte Carlo method, Metropolis et~al. (\citeyear{MetropolisRosenbluthRosenbluthTellerTeller1953}) propose
a random walk modification of the $N$ particles. That is, for each
particle $i$ $(1\le i\le N)$, values
\[
x^\prime_i = x_i + \sigma\xi_{1i} \quad\mbox{and}\quad
y^\prime_i = y_i + \sigma\xi_{2i}
\]
are proposed, where both $\xi_{1i}$ and $\xi_{2i}$ are uniform
$\mathcal{U}(-1,1)$. The energy difference $\Delta E$
between the new configuration and the previous one is then computed and
the new configuration is accepted
with probability
%
\begin{equation}\label{eq:MHstep}
\min\{1, \exp( -\Delta E/kT )\},
\end{equation}
and otherwise the previous configuration is replicated, in the sense
that its counter is increased by one
in the final average of the $F(\theta_t)$'s over the $\tau$ moves of
the random walk, $1\le t\le\tau$. Note
that Metropolis et~al. (\citeyear{MetropolisRosenbluthRosenbluthTellerTeller1953}) move one
particle at a time, rather than moving all of them together, which
ma\-kes the initial
algorithm appear as a primitive kind of Gibbs sampler!

The authors of Metropolis et~al. (\citeyear{MetropolisRosenbluthRosenbluthTellerTeller1953})
demonstrate the validity of the
algorithm by first establishing irreducibility, which they call
\textit{ergodicity}, and second proving ergodicity,
that is, convergence to the stationary distribution. The second part is
obtained via a discretization of the
space: They first note that the proposal move is reversible, then
establish that $\exp\{ -E/kT \}$ is invariant.
The result is therefore proven in its full generality, minus the discretization.
The number of iterations of the Metropolis algorithm used in the paper
seems to be limited: 16 steps for burn-in
and 48 to 64 subsequent iterations, which required four to five hours
on the Los Alamos computer.

An interesting variation
is the \textit{Simulated Annealing} algorithm, developed
by Kirkpatrick, Gelatt and Vecchi (\citeyear{kirkpatrickgelattvecchi1983}), who connected optimization with {\it
annealing}, the cooling of a metal. Their variation is to allow
the temperature $T$ in (\ref{eq:MHstep}) to change as the algorithm
runs, according to a ``cooling schedule,''
and the Simulated Annealing algorithm can be shown to find the global
maximum with probability $1$, although the
analysis is quite complex due to the fact that, with varying $T$, the
algorithm is no longer a~time-homogeneous Markov\vadjust{\goodbreak} chain.


\setattribute{citeyear}{font}{\sffamily}
\subsection{\texorpdfstring{The Hastings (\protect\citeyear{Hastings1970}) Paper}{The Hastings (1970) Paper}}
\setattribute{citeyear}{font}{\relax}

The Metropolis algorithm was later generalized by Hastings (\citeyear{Hastings1970}) and
his student Peskun (\citeyear{Peskun1973}, \citeyear{Peskun1981}) as a
statistical simulation tool that could overcome the curse of
dimensionality met by regular Monte Carlo methods,
a point already emphasized in Metropolis et~al. (\citeyear{MetropolisRosenbluthRosenbluthTellerTeller1953}).\footnote{In fact,
Hastings starts by mentioning
a decomposition of the target distribution into a \textit{product of
one-dimensional conditional distributions}, but this
falls short of an early Gibbs sampler.}

In his \textit{Biometrika} paper,\footnote{Hastings (\citeyear{Hastings1970}) is one of
the ten \textit{Biometrika} papers reproduced in
Titterington and Cox (\citeyear{titteringtoncox2001}).}
Hastings (\citeyear{Hastings1970}) also defines his methodology for finite and
reversible Mar\-kov chains, treating the continuous case
by using a discretization analogy. The generic probability of
acceptance for a move from state $i$ to state $j$ is
\[
\alpha_{ij} = \frac{s_{ij}}{1+(\pi_i/\pi_j)(q_{ij}/q_{ji})},
\]
where $s_{ij}=s_{ji}$, $\pi_i$ denotes the target and $q_{ij}$ the proposal.
This generic form of probability encompasses the forms of both
Metropolis et~al. (\citeyear{MetropolisRosenbluthRosenbluthTellerTeller1953})
and Barker (\citeyear{Barker1965}). At this stage,
Hastings mentions that \textit{little is known about the relative merits of
those two choices \textup{(even though)} Metropolis's method may be
preferable}. He also warns against \textit{high rejection
rates as indicative of a poor choice of transition matrix}, but does
not mention the opposite pitfall of low rejection
rates, associated with a slow exploration of the target.

The examples given in the paper are a Poisson~tar\-get with a $\pm1$
random walk proposal, a normal target with a uniform
random walk proposal mixed with its reflection, that is, a uniform
proposal centered at $-\theta_t$ rather than at the current
value $\theta_t$ of the Markov chain, and then a multivariate target
where Hastings introduces a~Gibbs sampling strategy, updating one
component at a~time and defining the composed transition
as satisfying the stationary condition because each component does
leave the target invariant. Hastings (\citeyear{Hastings1970}) actually
refers to Ehrman, Fosdick and Handscomb (\citeyear{ErhmanFosdickHandscomb1960}) as a preli\-minary, if
specific, instance of this sampler. More~pre\-cisely, this
is Metropolis-within-Gibbs except for the name.
This first introduction of the Gibbs sampler has thus
been completely overlooked, even though the proof of convergence is
completely general, based on a composition argument as
in Tierney (\citeyear{Tierney1994}),~dis\-cussed in Section \ref{sub:Adva}.
The remainder of the paper deals with (a) an importance sampling
version of MCMC, (b) general
remarks about assessment of~the error, and (c) an application to random
orthogonal matrices, with another example of
Gibbs sampling.

Three years later, 
Peskun (\citeyear{Peskun1973}) published a comparison of Metropolis' and Barker's
forms of acceptance probabilities and shows in a discrete setup that
the optimal choice
is that of Metropolis, where optimality is to be understood in terms of
the asymptotic variance of any empirical average. The proof is a direct
consequence of a result
by Kemeny and Snell (\citeyear{kemenysnell1960}) on the asymptotic variance. Peskun also
establishes that this asymptotic variance can
improve upon the i.i.d. case if and only if the eigenvalues of $\mathbf
{P}-\mathbf{A}$ are all negative, when $\mathbf{A}$ is the
transition matrix corresponding to i.i.d. simulation and $\mathbf{P}$ the
transition matrix corresponding to the
Metropolis algorithm, but he concludes that the trace of $\mathbf
{P}-\mathbf{A}$ is always positive.

\vspace*{-1pt}
\section{Seeds of the Revolution}\label{sec:gib}
\vspace*{-1pt}

A number of earlier pioneers had brought forward the seeds of Gibbs
sampling; in particular, Hammersley
and Clifford had produced a constructive argument in 1970 to recover a
joint distribution from its conditionals,
a result later called the \textit{Hammersley--Clifford} theorem by
Besag (\citeyear{Besag1974}, \citeyear{Besag1986}). Besides Has\-tings (\citeyear{Hastings1970})
and Geman and Geman (\citeyear{GemanGeman1984}),
already mentioned, other papers that contained the seeds of Gibbs
sampling are Besag and Clifford (\citeyear{BesagClifford1989}),
Broniatowski, Celeux and Diebolt (\citeyear{BroniatowskiCeleuxDiebolt1984}), Qian and Titterington (\citeyear{QianTitterington1990}) and
Tanner and Wong (\citeyear{TannerWong1987}).

\subsection{Besag's Early Work and the Fundamental (Missing) Theorem}
In the early 1970's, Hammersley, Clifford and Besag were working on the
specification of joint distributions
from conditional distributions and on necessary and sufficient
conditions for the conditional distributions to
be compatible with a joint distribution. What is now known as the
\textit{Hammersley--Clifford} theorem states that
a joint distribution for a vector associated with a~dependence graph
(edge meaning dependence and absence of
edge conditional independence) must be represented as a product of
functions over the \textit{cliques} of the graphs, that is,
of functions depending only on the components indexed by the labels in
the clique.\footnote{A clique is a
maximal subset of the nodes of a graphs such that every pair of nodes
within the clique is connected by an edge
(Cressie, \citeyear{Cressie1993}).}

From a historical point of view,
Hammersley (\citeyear{hammersley1974}) explains why the Hammersley--Clifford theorem was
never published as such, but only
through Besag (\citeyear{Besag1974}). The reason is that Clifford and Hammersley
were dissatisfied with the positivity constraint:
The joint density could be recovered from the full conditionals only
when the support of the joint was made of
the product of the supports of the full conditionals.
While they strived \textit{to make the theorem independent of any
positivity condition}, their graduate student published a
counter-example that put a full stop to their endeavors (Moussouris, \citeyear{moussouris1974}).

While Besag (\citeyear{Besag1974}) can certainly be credited to some extent of the
(re-)discovery of the
Gibbs sampler, Besag (\citeyear{Besag1975}) expressed doubt about the practicality
of his method, noting that ``the technique is unlikely to be
particularly helpful in many other than binary situations and the
Markov chain itself has no
practical interpretation,'' clearly understating the importance of his
own work.

%
%

A more optimistic sentiment was expressed earlier by
Hammersley and Handscomb (\citeyear{hammersleyhandscomb1964}) in their textbook on Monte Carlo methods.
There they cover such topics as ``Crude Monte Carlo,''
importance sampling, control variates and ``Conditional Monte Carlo,''
which looks surprisingly like a missing-data completion approach. Of
course, they do not cover the Hammersley--Clifford theorem, but
they state in the Preface:
\begin{quote}
We are convinced nevertheless that Monte Carlo methods will one day
reach an impressive maturity.
\end{quote}
Well said!

\subsection{EM and Its Simulated Versions as Precursors}
Because of its use for missing data problems, the EM algorithm (Dempster,
Laird and Rubin, \citeyear{dempsterlairdrubin1977})
has early connections with Gibbs sampling. For instance,
Broniatowski, Celeux and Diebolt (\citeyear{BroniatowskiCeleuxDiebolt1984}) and Celeux and Diebolt (\citeyear{CeleuxDiebolt1985}) had
tried to overcome the
dependence of EM methods on the starting value
by replacing the E step with a \textit{simulation}\vadjust{\goodbreak} step,
the missing data~$z$ being generated conditionally on the observation~$x$ and on the current value of the parameter $\theta_m$. The maximization
in the M step is then done on the simulated complete-data
log-likelihood, a predecessor to the Gibbs step of Diebolt and Ro\-bert (\citeyear{dieboltrobert1994}) for mixture estimation.
Unfortunately, the theoretical convergence results for these methods
are limited. Celeux and Diebolt (\citeyear{CeleuxDiebolt1990}) have, however,
solved the convergence problem of SEM\vadjust{\goodbreak} by devising a hybrid version
called SAEM (for \textit{Simulated Annealing EM}),
where the amount of randomness in the simulations decreases with the
iterations, ending up with an EM algorithm.\footnote{Other and more
well-known connections between EM and MCMC algorithms can be found in
the literature (Liu and Rubin, \citeyear{liurubin1994}; Meng and Rubin,
\citeyear{mengrubin1992}; Wei and Tanner, \citeyear{weitanner1990a}), but the connection with Gibbs sampling is
more tenuous in that the
simulation methods are used to approximate quantities in a Monte Carlo fashion.}

\vspace*{2pt}\subsection{Gibbs and Beyond}\vspace*{2pt}
Although somewhat removed from statistical inference in the classical sense
and based on earlier techniques used in Statistical Physics, the
landmark paper by Geman and Geman (\citeyear{GemanGeman1984})
brought Gibbs sampling into the arena of statistical application.
This paper is also responsible for the name \textit{Gibbs sampling},
because it implemented
this method for the Bayesian study of \textit{Gibbs random fields} which,
in turn, derive
their name from the physicist Josiah Willard Gibbs (1839--1903). This
original implementation
of the Gibbs sampler was applied to a discrete image processing problem
and did not involve
completion. But this was one more spark that led to the explosion, as
it had a clear influence
on Green, Smith, Spiegelhalter and others.

The extent to which Gibbs sampling and Metropolis algorithms were in
use within the image analysis
and point process communities is actually quite large, as illustrated
in Ripley (\citeyear{ripley1987}) where
Sec-\break tion~4.7 is entitled ``Metropolis' method and random fields'' and
describes the implementation
and the validation of the Metropolis algorithm in a finite setting with
an application to Markov
random fields and the corresponding issue of bypassing the normalizing
constant. Besag, York and Molli{\'e} (\citeyear{besagyorkmollie1991})
is another striking example of the activity in the spatial statistics
community at the end of the 1980s.

\vspace*{2pt}\section{The Revolution}\vspace*{2pt}

The gap of more than 30 years between Metropolis et~al. (\citeyear{MetropolisRosenbluthRosenbluthTellerTeller1953})
and Gelfand and Smith (\citeyear{GelfandSmith1990}) can still be partially attributed to
the lack of appropriate computing power, as most of the examples now
processed by
MCMC algorithms could not have been treated previously, even though the
hundreds of dimensions processed
in Metropolis et~al. (\citeyear{MetropolisRosenbluthRosenbluthTellerTeller1953}) were quite
formidable. However,
by the mid-1980s, the pieces were all in place.

After Peskun, MCMC in the statistical world was dormant for about 10 years,
and then several papers appeared that highlighted its usefulness in
specific settings
like pattern recognition, image analysis or spatial statistics.
In particular, Geman and Geman (\citeyear{GemanGeman1984})
influenced Gelfand and Smith (\citeyear{GelfandSmith1990}) to write a paper that is the
genuine starting point for an intensive
use of MCMC methods by the mainstream statistical community. It sparked
new interest in Bayesian
methods, statistical computing,
algorithms and stochastic processes through the use of computing
algorithms such as the Gibbs sampler and the
Metropolis--Hastings algorithm.
(See Casella and George, \citeyear{CasellaGeorge1992}, for an elementary introduction to the Gibbs
sampler.\footnote{On a humorous note, the original Technical Report of
this paper was called \textit{Gibbs for Kids}, which was changed because
a~referee did not appreciate the humor. However, our colleague Dan
Gianola, an Animal Breeder at Wisconsin, liked the title. In using
Gibbs sampling in his work, he gave a presentation in 1993 at the 44th
Annual Meeting of the European Association for Animal Production,
Arhus, Denmark. The title: \textit{Gibbs for Pigs}.})

Interestingly, the earlier paper by Tanner and\break Wong (\citeyear{TannerWong1987}) had
essentially the same ingredients as Gelfand and Smith (\citeyear{GelfandSmith1990}),
namely, the fact that simulating from the conditional distributions is
sufficient to asymptotically simulate from
the joint. This paper was considered important enough to be a discussion
paper in the \textit{Journal of the American Statistical Association}, but
its impact was somehow limited, compared with Gelfand and Smith (\citeyear{GelfandSmith1990}).
There are several reasons for this; one being that the method
seemed to only apply to missing data problems, this impression being
reinforced by
the name \textit{data augmentation}, and another is that the authors were
more focused on approximating the
posterior distribution. They suggested a MCMC approximation to the
target $\pi(\theta|x)$ at each
iteration of the sampler, based on
\begin{eqnarray*}
&&\frac{1}{m}\sum_{k=1}^m \pi(\theta|x,z^{t,k}),\\
&& z^{t,k}\sim\hat\pi
_{t-1}(z|x),\quad k=1, \ldots,m,
\end{eqnarray*}
that is, by replicating $m$ times the simulations from the current
approximation $\hat\pi_{t-1}(z|x)$ of the marginal posterior
distribution of the missing data. This focus on estimation of the
posterior distribution connected the original Data Augmentation
algorithm~to EM, as pointed out by Dempster in the discussion. Although
the discussion by Carl Morris gets very close to the two-stage Gibbs
sampler for hierarchical models, he is still concerned about doing $m$
iterations, and worries about how costly that would be. Tanner and Wong
mention taking $m=1$ at the end of the paper, referring to this as an
``extreme case.''

In a sense, Tanner and Wong (\citeyear{TannerWong1987}) were still too close to
Rubin's \citeyear{Rubin1978}
multiple imputation to start a~new revolution. Yet another reason for
this may be that the theoretical background was based on functional analysis
rather than Markov chain theory, which needed, in particular, for the
Markov kernel to be uniformly bounded
and equicontinuous. This may have discouraged potential users as
requiring too much mathematics.

The authors of this review were fortunate enough to attend many focused
conferences during this time, where we were
able to witness the explosion of Gibbs sampling. In the summer of 1986
in Bowling Green, Ohio, Adrian Smith gave
a series of ten lectures on hierarchical models. Although there was a
lot of computing mentioned, the Gibbs sampler
was not fully developed yet. In another lecture in June 1989 at a
Bayesian workshop in Sherbrooke,
Qu\'{e}bec, he revealed for the first time the generic features of the
Gibbs sampler, and
we still remember vividly the shock induced on ourselves and on the
whole audience by the sheer breadth of the method:
This development of Gibbs sampling, MCMC, and the resulting seminal
paper of Gelfand and Smith (\citeyear{GelfandSmith1990})
was an \textit{epiphany} in the world of Statistics.

\begin{definition*}[(Epiphany $n$)] A
spiritual event in which the essence of a
given object of manifestation appears to the subject, as in a sudden
flash of recognition.
\end{definition*}

The explosion had begun, and just two years later, at an MCMC
conference at Ohio State University
organized by Alan Gelfand, Prem Goel and Adrian Smith, there were three
full days of talks. The presenters at
the conference read like a Who's Who of MCMC, and the level, intensity
and impact of that conference, and the
subsequent research, are immeasurable. Many of the talks were to become
influential papers, including Albert and Chib (\citeyear{albertchib1993b}),
Gelman and Rubin (\citeyear{gelmanrubin1992}), Geyer (\citeyear{geyer1992}),
Gilks (\citeyear{gilks1992}), Liu, Wong and Kong
(\citeyear{liuwongkong1994}, \citeyear{liuwongkong1995}) and Tierney (\citeyear{Tierney1994}). The program
of the conference is reproduced in the \hyperref[app:ohio]{Appendix}.

Approximately one year later, in May of 1992, there was a meeting of
the Royal Statistical Society on ``The Gibbs
sampler and other Markov chain Monte Carlo methods,'' where four papers
were presented followed by much discussion.
The papers appear in the first volume of \textit{JRSSB} in 1993, together
with $49$(!) pages of discussion.
The excitement is clearly evident in the writings, even though the
theory and implementation were not always perfectly
understood.\footnote{On another humorous note, Peter Clifford opened
the discussion by noting ``$\ldots$we have had the opportunity to hear a
large amount about an important new area in statistics. It may well be
remembered as the `afternoon of the 11 Bayesians.' Bayesianism has
obviously come a long way. It used to be that you could tell a Bayesian
by his tendency to hold meetings in isolated parts of Spain and his
obsession with coherence, self-interrogation and other manifestations
of paranoia. Things have changed, and there may be a general lesson
here for statistics. Isolation is counter-productive.''}

Looking at these meetings, we can see the paths that Gibbs sampling
would lead us down. In the next two sections
we will summarize some of the advances from the early to mid 1990s.

\subsection{Advances in MCMC Theory}\label{sub:Adva}

Perhaps the most influential MCMC theory paper of the 1990s is Tierney (\citeyear{Tierney1994}), who carefully laid out
all of the assumptions needed to analyze the Markov chains and then
developed their properties, in particular,
convergence of ergodic averages and central limit theorems. In one of
the discussions of that paper,
Chan and Geyer (\citeyear{changeyer1994}) were able to relax a condition on Tierney's
Central Limit Theorem, and this new condition plays an
important role in research today (see Section \ref{sec:regeneration}).
A pair of very influential, and innovative,
papers is the work of Liu, Wong and Kong (\citeyear{liuwongkong1994}, \citeyear{liuwongkong1995}), who
very carefully analyzed the covariance
structure of Gibbs sampling, and were able to formally establish the
validity of Rao--Blackwellization in Gibbs
sampling. Gelfand and Smith (\citeyear{GelfandSmith1990}) had used Rao--Blackwellization, but
it was not justified at that time, as the
original theorem was only applicable to i.i.d. sampling, which is not the
case in MCMC. Another significant entry is
Rosenthal (\citeyear{rosenthal1995}), who obtained one of the earliest results on
exact rates of convergence.

Another paper must be singled out, namely, Men\-gersen and Tweedie (\citeyear{mengersentweedie1996}), for setting the tone for the study of
the speed of convergence of MCMC algorithms to the target distribution.
Subsequent works in this area by Richard Tweedie,
Gareth Ro\-berts, Jeff Rosenthal and co-authors are too numerous to be
mentioned here, even though the paper
by Roberts, Gelman and Gilks (\citeyear{robertsgelmangilks1997}) must be cited for setting explicit
targets on the acceptance rate of the
random walk Metropolis--Hastings algorithm, as well as Roberts and Rosenthal (\citeyear{robertsrosenthal1999}) for getting an upper bound
on the number of iterations $(523)$ needed to approximate the target up
to $1\%$ by a slice sampler. The untimely
death of Richard Tweedie in 2001, alas, had a major impact on the book
about MCMC convergence he was contemplating
with Gareth Roberts.

One pitfall arising from the widespread use of\break Gibbs sampling was the
tendency to specify models only through
their conditional distributions, almost always without referring to the
positivity conditions in Section~\ref{sec:gib}.
Unfortunately, it is possible to specify a~perfectly legitimate-looking
set of conditionals that do not correspond to
any joint distribution, and the resulting Gibbs chain cannot converge.
Hobert and Casella (\citeyear{hobertcasella1996}) were able to
document the conditions needed for a convergent Gibbs chain, and
alerted the Gibbs community to this problem,
which only arises when improper priors are used, but this is a frequent
occurrence.

Much other work followed, and continues to grow today. Geyer and Thompson (\citeyear{geyerthompson1995}) describe how to put a
``ladder'' of chains together to have both ``hot'' and ``cold'' exploration,
followed by Neal's \citeyear{neal1996} introduction of tempering;
Athreya, Doss and Sethuraman (\citeyear{athreyadosssethuraman1996}) gave more easily verifiable
conditions for convergence;
Meng and van Dyk (\citeyear{mengvandyk1999}) and Liu and Wu (\citeyear{liuwu1999}) developed the theory of
parameter expansion
in the Data Augmentation algorithm, leading to construction of chains
with faster convergence,
and to the work of Hobert and Marchev (\citeyear{hobertmarchev2008}), who give precise
constructions and theorems to
show how parameter expansion can uniformly improve over the original chain.

\vspace*{3pt}\subsection{Advances in MCMC Applications}\vspace*{3pt}

The real reason for the explosion of MCMC methods was the fact that an
enormous number of problems that were
deemed to be computational nightmares now cracked open like eggs. As an
example, consider this very simple random
effects model from Gelfand and Smith (\citeyear{GelfandSmith1990}). Observe
%
\begin{equation}\label{eq:mixed}
Y_{ij} = \theta_i + \varepsilon_{ij},\quad i=1, \ldots, K,\  j=1, \ldots, J,
\end{equation}
where
\begin{eqnarray*}
\theta_{i} &\sim& N(\mu, \sigma^2_\theta),\\
\varepsilon_{ij}&\sim& N(0, \sigma^2_\varepsilon),\quad \mbox{independent of $\theta_i$}.
\end{eqnarray*}
Estimation of the variance components can be difficult for a
frequentist (REML is typically preferred), but
it indeed was a nightmare for a Bayesian, as the integrals were
intractable. However, with the usual priors
on $\mu, \sigma^2_\theta$ and $ \sigma^2_\varepsilon$, the full
conditionals are trivial to sample from
and the problem is easily solved via Gibbs sampling. Moreover, we can
increase the number of variance components and the Gibbs solution
remains easy to implement.

During the early 1990s, researchers found that Gibbs, or
Metropolis--Hastings, algorithms would be able to give solutions to
almost any problem
that they looked at, and there was a veritable flood of papers applying
MCMC to previously intractable
models, and getting good answers. For example, building on (\ref
{eq:mixed}), it was quickly realized that
Gibbs sampling was an easy route to getting estimates in the linear
mixed models
(Wang, Rutledge and Gianola, \citeyear{wangrutledgegianola1993}, \citeyear{wangrutledgegianola1994}), and even
generalized linear mixed models
(Zeger and Karim, \citeyear{zegerkarim1991}).
Building on the experience gained with the EM algorithm, similar
arguments made it possible to analyze probit models using a
latent variable approach in a linear mixed model (Albert and Chib, \citeyear{albertchib1993b}),
and in mixture models with Gibbs sampling (Diebolt and Robert, \citeyear{dieboltrobert1994}).
It progressively dawned on the community that latent variables could be
artificially
introduced to run the Gibbs sampler in about every situation, as
eventually published in
Damien, Wakefield and Walker (\citeyear{damienwakefieldwalker1999}), the main example being the slice
sampler (Neal, \citeyear{neal2003}).
A very incomplete list of some other applications include changepoint analysis
(Carlin, Gelfand and Smith, \citeyear{carlingelfandsmith1992}; Stephens, \citeyear{stephens1994}), Genomics
(Churchill, \citeyear{churchill1995}; Lawrence et~al., \citeyear{lawaltbog1993}; Stephens and Smith, \citeyear{stephenssmith1993}),
capture--recapture (Dupuis, \citeyear{dupuis1995b}; George and Robert, \citeyear{georgerobert1992}), variable
selection in regression
(George and McCulloch, \citeyear{georgemcculloch1993}), spatial sta\-tistics (Raftery and Banfield, \citeyear{rafterybanfield1991}), and
longitudinal studies (Lange, Carlin and Gelfand, \citeyear{langecarlingelfand1992}).

Many of these applications were advanced though other developments such
as the
Adaptive Rejection Sampling of Gilks (\citeyear{gilks1992}); Gilks, Best and Tan (\citeyear{gilksbesttan1995}), and
the simulated tempering approaches of Geyer and Thompson (\citeyear{geyerthompson1995}) or
Neal (\citeyear{neal1996}).

\section{After the Revolution}

After the revolution comes the ``second'' revolution, but now we have a
more mature field.
The revolution has slowed, and the problems are being solved in,
perhaps, deeper and more
sophisticated ways, even though Gibbs\vadjust{\goodbreak} sampling also offers to the
amateur the
possibility to handle Bayesian analysis in complex models at little
cost, as exhibited by
the widespread use of BUGS, which mostly focuses\footnote{BUGS now
uses both Gibbs sampling and Metropolis--Hastings algorithms.} on this approach.
But, as before, the methodology continues to expand the set of problems
for which statisticians can provide meaningful solutions, and thus
continues to further the impact of Statistics.

\subsection{A Brief Glimpse at Particle Systems}\label{no:briefpart}

The realization of the possibilities of iterating importance sampling
is not new: in fact,
it is about as old as Monte Carlo methods themselves. It can be found
in the molecular
simulation literature of the 50s, as in Hammersley and Morton (\citeyear{HammersleyMorton1954}),
Rosenbluth and Rosenbluth (\citeyear{RosenbluthRosenbluth1955}) and Marshall (\citeyear{Marshall1965}).
Hammersley and colleagues proposed such a method to simulate a self-avoiding
random walk (see Madras and Slade, \citeyear{MadrasSlade1993}) on a grid, due to huge
inefficiencies in
regular importance sampling and rejection techniques.
Although this early implementation occurred in particle physics,
the use of the term ``particle'' only dates back to Kitagawa (\citeyear{Kitagawa1996}), while
Carpenter, Clifford and Fernhead (\citeyear{CarpenterCliffordFernhead1997}) coined the term ``particle filter.''
In signal processing, early occurrences of a particle filter can be
traced back to Handschin and Mayne (\citeyear{HandschinMayne1969}).

More in connection with our theme, the landmark paper of Gordon, Salmond and Smith (\citeyear{gordonsalmonsmith1993}) introduced
the bootstrap filter which, while formally connected with importance
sampling, involves
past simulations and possible MCMC steps (Gilks and Berzuini, \citeyear{gilksberzuini2001}). As
described in
the volume edited by Doucet, de Freitas and Gordon (\citeyear{doucetdefreitasgordon2001}), particle
filters are simulation methods
adapted to sequential settings where data are collected progressively
in time, as in radar detection,
telecommunication correction or financial volatility estimation. Taking
advantage of state-space
representations of those dynamic models, particle filter methods
produce Monte Carlo approximations
to the posterior distributions by propagating simulated samples who\-se
weights are actualized against
the incoming observations. Since the importance weights have a tendency
to degenerate, that is, all weights
but one are close to zero, additional MCMC steps can be introduced at
times to recover the variety
and representativeness of the sample. Modern connections with MCMC in
the construction of the proposal kernel
are to be found, for instance, in Doucet, Godsill and Andrieu (\citeyear{doucetgodsillandrieu2001}) and
in Del~Moral, Doucet and Jasra (\citeyear{delmoraldoucetjasra2006}). At the same time, sequential
imputation was developed in Kong, Liu and Wong (\citeyear{KongLiuWong1994}), while
Liu and Chen (\citeyear{LiuChen1995}) first formally pointed out the importance of
resampling in
sequential Monte Carlo, a~term coined by them.

The recent literature on the topic more closely bridges the gap between
sequential Monte Carlo
and MCMC methods by making adaptive MCMC a possibility (see, e.g.,
Andrieu et~al., \citeyear{andrieudefreitasdoucetjordan2004}, or Roberts and Rosenthal, \citeyear{robertsrosenthal2005}).

\subsection{Perfect Sampling}\label{sec:cftistory}
Introduced in the seminal paper of Propp and Wilson (\citeyear{ProppWilson1996}),
perfect sampling, namely, the ability to use MCMC methods to produce an
exact (or perfect) simulation
from the target, maintains a unique place in the history of MCMC
methods. Although this exciting discovery led
to an outburst of papers, in particular, in the large body of work of
M\o ller
and coauthors, including the book by M{{\o}}ller and Waagepetersen (\citeyear{MoellerWaagepetersen2003}),
as well
as many reviews and introductory materials, like Casella, Lavine and Robert (\citeyear{CasellaLavineRobert2001}),
Fismen (\citeyear{Fismen1998}) and Dimakos (\citeyear{Dimakos2001}), the excitement quickly
dried out. The major
reason for this ephemeral lifespan is that the construction of perfect
samplers is most often close to
impossible or impractical,
despite some advances in the implementation (Fill, \citeyear{fill1998a}, \citeyear{fill1998b}).

There is, however, ongoing activity in the area of point
processes and stochastic geometry, much from the work of M\o ller
and Kendall. In particular, Kendall and M{\o}ller (\citeyear{KendallMoeller2000}) developed an alternative
to the \textit{Coupling From The Past} (CFPT) algorithm of Propp and Wilson (\citeyear{ProppWilson1996}), called \textit{horizontal CFTP},
which mainly applies to point processes and is based on continuous time
birth-and-death processes.
See also Fern{\'a}ndez, Ferrari and Garcia (\citeyear{FernandezFerrariGarcia1999}) for
another horizontal CFTP algorithm for point processes. Berthelsen and M{{\o}}ller (\citeyear{BerthelsenMoeller2003})
exhibited a use of these algorithms for nonparametric Bayesian
inference on point processes.

\subsection{Reversible Jump and Variable Dimensions}

From many viewpoints, the invention of the reversible jump algorithm in
Green (\citeyear{green1995}) can be seen as the
start of the second MCMC revolution: the formalization of a Markov
chain that moves across models and parameter spaces
allowed for the Baye\-sian processing of a wide variety of new models and
contributed to the success of Bayesian
model choice and subsequently to its adoption in other fields. The\-re
exist earlier alternative Monte
Carlo solutions like Gelfand and Dey (\citeyear{gelfanddey1994}) and Carlin and Chib (\citeyear{carlinchib1995}),
the later being very close in
spirit to reversible jump MCMC (as shown by the completion scheme of
Brooks, Giudici and Roberts, \citeyear{brooksgiudiciroberts2003}),
but the definition of a proper
balance condition on cross-model Markov kernels in Green (\citeyear{green1995})
gives a generic setup for exploring
variable dimension spaces, even when the number of models under
comparison is infinite. The impact of this
new idea was clearly perceived when looking at the First European
Conference on Highly Structured Stochastic Systems that
took place in Rebild, Denmark, the next year, organized by Stephen
Lauritzen and Jesper M\o ller: a large majority
of the talks were aimed at direct implementations of RJMCMC to various
inference problems. The application of RJMCMC to
mixture order estimation in the discussion paper of Richardson and Green (\citeyear{richardsongreen1997}) ensured further dissemination of the
technique. Continuing to develop RJMCMCt, Stephens (\citeyear{stephens2000}) proposed
a continuous time version of RJMCMC, based on earlier ideas
of Geyer and M{\o}ller (\citeyear{geyermoller1994}), but with similar
properties (Capp{\'e}, Robert and Ryd{\'e}n, \citeyear{capperobertryden2003}),
while Brooks, Giudici and Roberts (\citeyear{brooksgiudiciroberts2003})
made proposals for increasing the efficiency of the moves. In
retrospect, while reversible jump is somehow unavoidable in
the processing of very large numbers of models under comparison, as,
for instance, in variable selection (Marin and Robert, \citeyear{marinrobert2007}),
the implementation of a complex algorithm like RJMCMC for the
comparison of a few models is somewhat of an overkill since there
may exist alternative solutions based on model specific MCMC chains,
for example (Chen, Shao and Ibra\-him, \citeyear{chenshaoibrahim2000}).

\subsection{Regeneration and the CLT}\label{sec:regeneration}
While the Central Limit Theorem (CLT) is a central tool in Monte Carlo
convergence assessment, its
use in MCMC setups took longer to emerge, despite early signals by
Geyer (\citeyear{geyer1992}),
and it is only recently that sufficiently clear conditions
emerged. We recall that the Ergodic Theorem (see, e.g., Robert and Casella, \citeyear{robertcasella2004}, Theorem 6.63)
states that, if $(\theta_t)_t$ is a Markov
chain with stationary distribution $\pi$, and $h(\cdot)$ is a
function with finite variance, then under fairly mild conditions,
%
\begin{equation}\label{eq:ergo}
\lim_{n \rightarrow\infty} \bar h_n = \int h(\theta) \pi(\theta)\,d\theta= \mathrm{E}_\pi h(\theta),
\end{equation}
almost everywhere, where $\bar h_n=(1/n)\sum_{i=1}^n
h(\theta_i)$.\break
For the CLT to be used to monitor this convergence,
%
\begin{equation}\label{eq:CLT}
\frac{\sqrt{n}(\bar h_n-\mathrm{E}_\pi h(\theta))}{\sqrt{\operatorname{Var}
h(\theta)}} \rightarrow\mathrm{N}(0,1),
\end{equation}
there are two roadblocks. First, convergence to normality is strongly
affected by the lack of independence.
To get CLTs for Markov chains, we can use a result of Kipnis and Varadhan (\citeyear{kipnisvaradhan1986}), which requires the chain to be
reversible, as is the case for holds for Metropolis--Hastings chains,
or we must delve into mixing conditions (Billingsley, \citeyear{billingsley1995},
Section 27),
which are typically not easy to verify. However, Chan and Geyer (\citeyear{changeyer1994})
showed how the condition of geometric ergodicity
could be used to establish CLTs for Markov chains. But getting the
convergence is only half of the problem. In order to
use~(\ref{eq:CLT}), we must be able to consistently estimate the
variance, which turns out to be another difficult endeavor.
The ``na\"{\i}ve'' estimate of the usual standard error is not
consistent in the dependent case
and the most promising paths for consistent variance estimates seems to
be through regeneration and batch means.

The theory of regeneration uses the concept of a~split chain (Athreya and Ney, \citeyear{athreyaney1978}),
and allows us to independently restart the chain
while preserving the stationary distribution. These independent
``tours'' then allow the calculation
of consistent variance estimates and honest monitoring of convergence
through (\ref{eq:CLT}). Early work
on applying regeneration to MCMC chains was done by Mykland, Tierney and Yu (\citeyear{myklandtierneyyu1995}) and Robert (\citeyear{robert1995b}),
who showed how to construct the chains and use them for variance
calculations and diagnostics (see also
Guihenneuc-Jouyaux and Robert, \citeyear{guihenneucjouyauxrobert1998}),
as well as deriving adaptive MCMC algorithms (Gilks, Roberts and Sahu, \citeyear{gilksrobertssahu1998}). Rosenthal (\citeyear{rosenthal1995})
also showed how to construct and use regenerative chains, and much of
this work is reviewed in Jones and Hobert (\citeyear{joneshobert2001}).
The most interesting and practical developments, however, are in Hobert et~al. (\citeyear{hobertjonespresnelrosenthal2002})
and Jones et~al. (\citeyear{jonesharancaffoneath2006}), where consistent estimators are
constructed for $\operatorname{Var}h(X)$, allowing
valid monitoring of convergence in chains that satisfy the CLT.
Interestingly, although
Hobert et~al. (\citeyear{hobertjonespresnelrosenthal2002}) use regeneration, Jones et~al. (\citeyear{jonesharancaffoneath2006})
get their consistent estimators thorough another technique, that of
cumulative batch means.

\section{Conclusion}

The impact of Gibbs sampling and MCMC was~to
change our entire method of thinking and\vadjust{\goodbreak} \mbox{attacking} problems,
representing a \textit{paradigm shift}
(Kuhn, \citeyear{kuhn1996}). Now, the collection
of real problems that we could solve grew almost without bound. Markov
chain Mon\-te Carlo changed our emphasis from ``closed
form'' solutions to algorithms, expanded our impact to solving ``real''
applied problems and to improving
numerical algorithms using statistical ideas, and led us into a world
where ``exact'' now means ``simulated.''

This has truly been a quantum leap in the evolution of the field of
statistics, and the evidence is that there are no
signs of slowing down. Although the ``explosion'' is over, the current
work is going deeper into theory and applications,
and continues to expand our horizons and influence by increasing our
ability to solve even bigger and more important
problems. The size of the data sets, and of the models, for example, in
genomics or climatology, is something that could
not have been conceived $60$ years ago, when Ulam and von Neumann
invented~the Monte Carlo method. Now we continue to plod on,
and hope that the advances that we make here will, in some way, help
our colleagues $60$ years in the future solve the
problems that we cannot yet conceive.

\appendix
\section*{Appendix: Workshop on Bayesian~Computation}\label{app:ohio}

This section contains the program of the Workshop on \textit{Bayesian
Computation via Stochastic Simulation}, held at Ohio State University,
February 15--17, 1991. The organizers, and their affiliations at the
time, were Alan Gelfand, University of Connecticut, Prem Goel, Ohio
State University, and Adrian Smith, Imperial College, London.

\begin{itemize}
\item \textit{Friday, Feb. 15, 1991.}
\begin{enumerate}[(a)]
\item[(a)] Theoretical Aspect of Iterative Sampling,\break Chair: Adrian
Smith.
\begin{longlist}
\item[(1)] Martin Tanner, University of Rochester: \textit{EM, MCEM, DA
and PMDA}.
\item[(2)] Nick Polson, Carnegie Mellon University: \textit{On the
Convergence of the Gibbs Sampler and Its Rate}.
\item[(3)] Wing-Hung Wong, Augustin Kong and Jun Liu, University of
Chicago: \textit{Correlation Structure and Convergence of the
Gibbs
Sampler and Related Algorithms}.
\end{longlist}
\item[(b)] Applications---I, Chair: Prem Goel.
\begin{longlist}
\item[(1)] Nick Lange, Brown University, Brad Carlin, Carnegie Mellon
University and Alan Gelfand, University of Connecticut:
\textit{Hierarchical Bayes Models for Progression of HIV Infection}.
\item[(2)] Cliff Litton, Nottingham University,\break England:
\textit{Archaeological Applications of Gibbs Sampling}.
\item[(3)] Jonas Mockus, Lithuanian Academy of Sciences, Vilnius:
\textit{Bayesian Approach to Global and Stochastic Optimization}.
\end{longlist}
\end{enumerate}
\item \textit{Saturday, Feb. 16, 1991.}
\begin{enumerate}[(a)]
\item[(a)] Posterior Simulation and Mar\-kov Sampling, Chair: Alan
Gelfand.
\begin{longlist}
\item[(1)] Luke Tierney, University of Minnesota: \textit{Exploring
Posterior Distributions Using Mar\-kov Chains}.
\item[(2)] Peter Mueller, Purdue University:\break
 \textit{A~Ge\-neric Approach
to Posterior Integration and Bayesian Sampling}.
\item[(3)] Andrew Gelman, University of California, Berkeley and
Donald P. Rubin, Harvard University: \textit{On the Routine Use of Markov
Chains for Simulations}.
\item[(4)] Jon Wakefield, Imperial College, London:
\textit{Parameterization Issues in Gibbs Sampling}.
\item[(5)] Panickos Palettas, Virginia Polytechnic Institute:
\textit{Acceptance--Rejection Method in Posterior Computations}.
\end{longlist}
\item[(b)] Applications---II, Chair: Mark Berliner.
\begin{longlist}
\item[(1)] David Stephens, Imperial College, London: \textit{Gene
Mapping via Gibbs Sampling}.
\item[(2)] Constantine Gatsonis, Harvard University: \textit{Random
Efleeds Model for Ordinal Cateqorica! Data with an Application to ROC
Analysis}.
\item[(3)] Arnold Zellner, University of Chicago, Luc Bauwens and
Herman Van Dijk: \textit{Bayesian Specification Analysis and Estimation
of~Simul\-taneous Equation Models Using Monte Carlo Methods}.
\end{longlist}
\item[(c)] Adaptive Sampling, Chair: Carl Morris.
\begin{longlist}
\item[(1)] Mike Evans, University of Toronto and Carnegie Mellon
University: \textit{Some Uses of\break Adaptive Importance Sampling and
Chaining}.
\item[(2)] Wally Gilks, Medical Research Council, Cambridge, England:
\textit{Adaptive Rejection Sampling}.
\item[(3)] Mike West, Duke University: \textit{Mixture Model
Approximations, Sequential Updating and Dynamic Models}.
\end{longlist}
\end{enumerate}
\item \textit{Sunday, Feb. 17, 1991}.
\begin{enumerate}[(c)]
\item[(a)] Generalized Linear and Nonlinear Models,\break Chair: Rob Kass.
\begin{longlist}
\item[(1)] Ruey Tsay and Robert McCulloch, University of Chicago:
\textit{Bayesian Analysis of Autoregressive Time Series}.
\item[(2)] Christian Ritter, University of Wisconsin: \textit{Sampling
Based Inference in Non Linear Regression}.
\item[(3)] William DuMouchel, BBN Software,\break Boston: \textit{Application
of the Gibbs Sampler to Variance Component Modeling}.
\item[(4)] James Albert, Bowling Green University and Sidhartha Chib,
Washington University, St. Louis: \textit{Bayesian Regression Analysis of
Binary Data}.
\item[(5)] Edwin Green and William Strawderman, Rutgers University:
\textit{Bayes Estimates for the Linear Model with Unequal Variances}.
\end{longlist}
\item[(b)] Maximum Likelihood and Weighted Boots\-trapping, Chair: George
Casella.
\begin{longlist}
\item[(1)] Adrian Raftery and Michael Newton, University of
Washington: \textit{Approximate Baye\-sian Inference by the Weighted
Bootstrap}.
\item[(2)] Charles Geyer, Universlty of Chicago: \textit{Monte Carlo
Maximum Likelihood via Gibbs Sampling}.
\item[(3)] Elizabeth Thompson, University of\break Washington:
\textit{Stochastic Simulation for Complex Genetic Analysis}.
\end{longlist}
\item[(c)] Panel Discussion---Future of Bayesian Inference Using Stochastic
Simulation, Chair: Prem Gael.
\begin{enumerate}
\item[$\bullet$] Panel---Jim Berger, Alan Gelfand and~Ad\-rian Smith.
\end{enumerate}
\end{enumerate}
\end{itemize}

\section*{Acknowledgments}
We are grateful for comments and suggestions\break from~Brad Carlin, Olivier Capp\'{e}, David Spiegelhalter, Alan Gelfand,
Peter Green, Jun Liu,
Sharon McGray\-ne, Peter M\"{u}ller, Gareth Roberts and Adrian Smith.
Christian Robert's work was partly done during a~vi\-sit to the
Queensland University of Technology,
Brisbane, and the author is grateful to Kerrie Mengersen for her
hospitality and support.
Supported by the Agence Nationale de la Recherche (ANR, 212, rue de
Bercy 75012 Paris)
through the 2006--2008 project ANR=05-BLAN-0299 \textsf{Adap'MC}. Supported
by NSF Grants DMS-06-31632, SES-06-31588 and MMS 1028329.

%

\end{document}